\documentstyle[epsfig,12pt]{article}
\newcommand{\bbb}{$B^0 \bar B^0$}

\def\greaterthansquiggle{\raise.3ex\hbox%
                        {$>$\kern-.75em\lower1ex\hbox{$\sim$}}}
\def\lessthansquiggle{\raise.3ex\hbox{$<$\kern-.75em\lower1ex\hbox{$\sim$}}}
\newcommand{\be}{\begin{equation}}
\newcommand{\ee}{\end{equation}}
\newcommand{\ba}{\begin{eqnarray}}
\newcommand{\ea}{\end{eqnarray}}
\newcommand{\no}{\nonumber}

\normalsize
\begin{document}
\bibliographystyle{plain}
\begin{titlepage}
\begin{flushright}
UWThPh-1997-34\\
October 1997\\
\end{flushright}
\vspace{15mm}
\begin{center}
{\Large \bf How devious are deviations from\\[5pt] 
quantum mechanics:
the case of the $B^0 \bar B^0$ system}\\[70pt]

R.A. Bertlmann and W. Grimus\\
Institut f\"ur Theoretische Physik\\
Universit\"at Wien\\
Boltzmanngasse 5\\
A-1090 Vienna, Austria

\vspace{3cm}

{\bf Abstract}\\[7pt]
\end{center}
Considering semileptonic decays of the entangled \bbb\ state which
is generated in the decay of
$\Upsilon(4S)$, we simply multiply the
quantum-mechanical interference term by a factor $(1-\zeta)$ and
use the ``decoherence parameter'' $\zeta$ as a measure for deviations from
quantum mechanics. We investigate several consequences of this
modification of semileptonic \bbb\ decays.
In particular, we show that when confronted with the experimental values of 
the ratio 
$R=\,$(\# like-sign dilepton events)/(\# opposite-sign dilepton events)
and of the $B_H$--$B_L$ mass difference, the ensuing
one standard deviation range of
the decoherence parameter depends strongly on the basis in the
$B^0$--$\bar B^0$ 
space used to build the entangled \bbb\ state. On the other
hand, in quantum mechanics physical quantities are, of course,
independent of such arbitrary basis choices.\\[14pt]

\end{titlepage}

\section{Introduction}

In recent years there is increasing interest in testing quantum
systems exhibiting Einstein--Podolsky--Rosen correlations. 
Such systems are suitable to discriminate between quantum mechanics 
and any local realistic (hidden variable) theory \cite{Bell}
(see, e.g., Ref. \cite{Bertlmann} for a short review). Usually
such experiments are carried out by using photons (see, e.g., Refs.
\cite{Aspect,Kwiat}), however, we find it desirable to perform
such tests also with massive particles. Appropriate quantum
systems are given by $K^0 \bar K^0$ \cite{Six,Selleri,ebe} and \bbb\
\cite{Datta,Kayser}. In particular, the entangled \bbb\ system
produced at the $\Upsilon (4S)$ offers the possibility to test
quantum-mechanical interference over macroscopic distances of
order $3 \times 10^{-2}$ mm.

In a previous work \cite{bg} we have studied
this entangled \bbb\ state by simply multiplying the
quantum-mechanical interference term by a factor $(1-\zeta)$ and
using the ``decoherence parameter'' $\zeta$ \cite{ebe} as a
measure for deviations from quantum mechanics. Confronting the ratio
$R=\,$(\# like-sign dilepton events)/(\# opposite-sign dilepton events)
with experimental values restricts the decoherence parameter to
$\zeta \leq 0.53$ (90\% CL). This is a result which conforms
nicely with quantum mechanics and shows that local realistic
theories $(\zeta = 1)$ are disfavoured.

Recently, Dass and Sarma \cite{ds} have performed a similar
analysis of the same entangled \bbb\ state, however, employing
the basis provided by the mass eigenstates $B_H$, $B_L$ rather
than the flavour states $B^0$, $\bar B^0$ which were in used in 
Ref. \cite{bg}.
When they confront the ratio $R$ with the data they obtain a
result which is very close to the expectation of quantum
mechanics, being nearly 8 standard deviations away from complete decoherence.
Of course, there is no contradiction between the results of Ref.
\cite{bg} and \cite{ds}.
Evidently, different basis choices in the $B^0$--$\bar B^0$
space produce in general different quantum-mechanical interference terms,
each with its own decoherence parameter.
When different interference terms get modified by decoherence parameters 
their values inferred from experimental input will also be
different in general.

In this paper we generalize the considerations of Refs.
\cite{bg,ds} by representing the \bbb\ state with the help of a general
basis in the $B^0$--$\bar B^0$ space. We will show that, given a basis, 
the ensuing range of the corresponding decoherence parameter 
depends strongly on that particular basis choice. In addition, arbitrary
basis choices can also mimic CP violation in the \bbb\ system.

\section{The formalism}

The decay $\Upsilon(4S) \to B^0 \bar B^0$ generates the state
\be \label{psi}
\Psi(t=0) = \frac{1}{\sqrt{2}} \left( |B^0 \rangle \otimes |\bar B^0 \rangle
-|\bar B^0 \rangle \otimes |B^0 \rangle \right)
\ee
with the charge conjugation quantum number $C=-1$. Eq.
(\ref{psi}) is the point of departure in Ref. \cite{bg}. 
On the other hand,
the mass eigenstates of the neutral $B$ mesons are given by
\be
| B_H \rangle = p |B^0 \rangle + q |\bar B^0 \rangle\, ,
\quad
| B_L \rangle = p |B^0 \rangle - q |\bar B^0 \rangle
\quad
\mathrm{with} \quad |p|^2 + |q|^2 = 1 \, .
\ee
Starting with this
basis, the state (\ref{psi}) is
rewritten as
\be \label{psi1}
\Psi(t=0) = -\frac{1}{2\sqrt{2}pq} \left( |B_H \rangle \otimes |B_L \rangle
-|B_L \rangle \otimes |B_H \rangle \right) \, .
\ee
It has been stressed in Ref. \cite{ds} that taking Eq.
(\ref{psi1}) and modifying the interference terms by $(1-\zeta)$
the results are different from those obtained in Ref.
\cite{bg} with the starting point Eq. (\ref{psi}). Of course,
with respect to quantum mechanics the states (\ref{psi}) and
(\ref{psi1}) are identical.

To generalize this consideration we take
an arbitrary basis
\be \label{basis}
|b_j \rangle = S_{1j} |B^0 \rangle + S_{2j} |\bar B^0 \rangle
\quad \mathrm{with} \quad j=1,2
\ee 
such that
\be \label{psiS}
\Psi(t=0) = \frac{1}{\sqrt{2} \det S} \left( |b_1 \rangle \otimes |b_2 \rangle
-|b_2 \rangle \otimes |b_1 \rangle \right) \, .
\ee
The discussions in Refs. \cite{bg} and \cite{ds} correspond to
the special cases $S = \mathbf{1}$ and $S = M$ with
\be
M = \left( \begin{array}{rr} p & p \\ q & -q \end{array} \right) \, ,
\ee
respectively.

The time evolution of the basis vectors (\ref{basis}) is given by
\be \label{bt}
|b_j(t) \rangle = (M \hat{g}(t) M^{-1}S)_{1j} |B^0 \rangle +
                  (M \hat{g}(t) M^{-1}S)_{2j} |\bar B^0 \rangle
\ee
with $\hat{g} = \mbox{diag}\, (g_H, g_L)$ where the functions
\be
g_{H,L}(t) = \exp \left( -i(m_{H,L} - \frac{i}{2} \Gamma_{H,L})t \right)
\ee
correspond to the time evolution of $|B_H \rangle$ and $|B_L
\rangle$, respectively. Eq. (\ref{bt}) can be simplified by
\be
M \hat{g}(t) M^{-1} = 
\left( \begin{array}{cc}
g_+(t) & \frac{p}{q} g_-(t) \\ \frac{q}{p} g_-(t) & g_+(t)
\end{array} \right)
\ee
with
\be
g_\pm(t) = \frac{1}{2} (g_H(t) \pm g_L(t)) \, .
\ee
Then we obtain the time evolution of the state (\ref{psi})
in terms of the basis (\ref{basis}) as \cite{car}
\be \label{tt}
|\Psi;t,t' \rangle = \frac{1}{\sqrt{2} \det S} 
(|b_1(t) \rangle \otimes |b_2(t') \rangle - 
|b_1(t) \rangle \otimes |b_2(t') \rangle ) 
\ee
or
\be \label{tt1}
|\Psi;t,t' \rangle = \frac{1}{\sqrt{2} \det S}
\left( T_{j1}(t) T_{k2}(t') - T_{j2}(t) T_{k1}(t') \right) 
|\beta_j \rangle \otimes |\beta_k \rangle
\ee
where a summation over equal indices is understood and we have defined
\be
|\beta_1 \rangle \equiv |B^0 \rangle, \quad 
|\beta_2 \rangle \equiv |\bar B^0 \rangle
\ee
and
\be
T(t) \equiv M \hat{g}(t) M^{-1} S \, .
\ee
Eq. (\ref{tt1}) follows from Eq. (\ref{tt}) by
inserting Eq. (\ref{bt}).

We want to underline the invariance of quantum mechanics under
the arbitrary basis choice (\ref{basis}) by formulating the
following theorem.
\newtheorem{theorem}{Theorem}
\begin{theorem}
The matrix element $\langle f_1 \otimes f_2 | \Psi;t,t' \rangle$ 
where $|f_1 \rangle$, $|f_2 \rangle$ are arbitrary states
is independent of
the matrix $S$ which characterizes the basis choice in the 
$B^0$--$\bar B^0$ space. 
\end{theorem}
\noindent \textbf{Proof:} The theorem follows from the fact that
\be
(M \hat{g}(t) M^{-1})_{j1} |\beta_j \rangle = |B^0(t) \rangle
\quad \mbox{and} \quad
(M \hat{g}(t) M^{-1})_{j2} |\beta_j \rangle = |\bar B^0(t) \rangle
\ee
describe the time evolutions of 
$|B^0 \rangle$ and $|\bar B^0 \rangle$, respectively,
such that we obtain
\be \label{inv}
\langle f_1 \otimes f_2 | \Psi;t,t' \rangle = \frac{1}{\sqrt{2}}
\left( 
\langle f_1|B^0(t) \rangle \langle f_2|\bar B^0(t') \rangle -
\langle f_1|\bar B^0(t) \rangle \langle f_2|B^0(t') \rangle \right)\, .
\ee
Evidently, $S$ does not occur on the right-hand side of Eq. (\ref{inv}).
\hspace*{\fill} $\Box$

\medskip

Concentrating now on inclusive semileptonic decays, we note that
$\ell^+$ tags $B^0$ whereas $\ell^-$ originates from a 
$\bar B^0$ decay. Defining
\be
b_+ = \sum_X | {\cal A}(B^0 \to X \ell^+ \nu_\ell) |^2
\quad \mbox{and} \quad
b_- = \sum_X | {\cal A}(\bar B^0 \to \bar X \ell^- \bar \nu_\ell) |^2
\ee
we obtain the following expressions for the number of dilepton events
\cite{car,pai,oku}:
\ba 
N_{++} & = & \frac{1}{2 \Gamma} b_+^2 \left| \frac{p}{q} \right|^2 I_-
+ \zeta \frac{1}{|\det S|^2} \, b_+^2 
\left| \int_0^\infty dt T_{11}^*(t) T_{12}(t) \right|^2, \label{++}\\
N_{--} & = & \frac{1}{2 \Gamma} b_-^2 \left| \frac{q}{p} \right|^2 I_-
+ \zeta \frac{1}{|\det S|^2} \, b_-^2 
\left| \int_0^\infty dt T_{22}^*(t) T_{21}(t) \right|^2, \label{--}\\
N_{+-} = N_{-+} & = & \frac{1}{2 \Gamma} b_+ b_- I_+ \quad 
       + \zeta \frac{1}{|\det S|^2} b_+ b_- \no \\ 
 & & \qquad \qquad \times \, \mbox{Re} 
       \left( \int_0^\infty dt \, T_{11}^*(t) T_{12}(t)
       \int_0^\infty dt' \, T_{22}^*(t') T_{21}(t') \right). \label{+-}
\ea
These equations show that deviations from quantum mechanics
parameterized by the decoherence parameter $\zeta$ are all
characterized by the following two integrals:
\ba
\lefteqn{\int_0^\infty dt \, T_{11}^*(t) T_{12}(t) = } \no \\
& & I_+ S_{11}^*S_{12} + \left| \frac{p}{q} \right|^2 
I_- S_{21}^*S_{22}
+ \frac{p}{q} I_{+-} S_{11}^*S_{22}
+ \left( \frac{p}{q} \right)^* I_{-+} S_{21}^*S_{12}, \label{int1}\\
\lefteqn{\int_0^\infty dt \, T_{22}^*(t) T_{21}(t) = } \no \\
& & I_+ S_{22}^*S_{21} + \left| \frac{q}{p} \right|^2 
I_- S_{12}^*S_{11}
+ \frac{q}{p} I_{+-} S_{22}^*S_{11}
+ \left( \frac{q}{p} \right)^* I_{-+} S_{12}^*S_{21}, \label{int2}
\ea
where
\ba
I_\pm & = & \int_0^\infty dt \, | g_\pm(t) |^2 = 
\frac{1}{2\Gamma} \left( \frac{1}{1-y^2} \pm \frac{1}{1+x^2}
\right) \, ,\no \\ 
I_{+-} = (I_{-+})^* & = & \int_0^\infty dt \, g_+(t)^* g_-(t) =
-\frac{1}{2\Gamma} \left( \frac{y}{1-y^2} + i\frac{x}{1+x^2}
\right) \, ,
\ea
and $x$ and $y$ are defined as
\be
x = \frac{\Delta m}{\Gamma} \quad \mbox{and} \quad 
y = \frac{\Delta \Gamma}{2 \Gamma}.
\ee
CPT invariance leads to $b_+ = b_-$ to a very good approximation.

As in our previous work Ref. \cite{bg}, we have introduced the decoherence
parameter $\zeta$ in Eqs. (\ref{++}), (\ref{--}) and (\ref{+-}) 
by multiplying with it the
interference terms which appear because $|\Psi;t,t' \rangle$ is
the sum of two terms (see Eqs. (\ref{tt}) and (\ref{tt1})). 
We want to stress once more that actually 
one should consider cases with different $S$ as different
extensions of the quantum-mechanical case and thus label $\zeta$
by $S$. This is also manifest from the fact that for
$\zeta \neq 0$ the expressions for the number of dilepton events depend on
$S$. We omit such a label for the sake of simplicity.

\section{Like-sign dilepton events and CP violation}

If quantum mechanics is valid, then CP violation in \bbb\ mixing
occurs for
$|p/q| \neq 1$ which is equivalent to $N_{++} \neq N_{--}$.
However, because of the introduction of the matrix $S$, which is
arbitrary apart from $\det S \neq 0$, even for $|p/q| = 1$ we
have $N_{++} \neq N_{--}$ in general\footnote{This can, e.g., be
seen by setting $S_{11} = S_{22} = 1$ and $S_{21} = 0$ and
varying $S_{12}$.}. This
is one of the strange consequences of the simple modification of
the quantum-mechanical expressions by the decoherence parameter:
the matrix $S$ mimics CP violation and the basis choice (\ref{basis})
which is unphysical at the level of quantum
mechanics gets a physical meaning because $N_{++}-N_{--}$
depends on $S$.

On the other hand, the cases $S=\mathbf{1}$ and $S=M$ discussed in
Refs. \cite{bg} and \cite{ds} lead
indeed to $N_{++} = N_{--}$ for $|p/q| = 1$ as can easily be
checked by inspecting Eqs. (\ref{int1}) and (\ref{int2}). 
It is difficult to give the general conditions for $S$ such that
in the limit $|p/q| \to 1$ the corresponding limit of
$N_{++}-N_{--}$ is zero. 

However, apart from the two special cases mentioned above, 
there are conditions such that this requirement is fulfilled for
a whole class of matrices $S$.
\begin{theorem}
If $y=0$ and the matrix $S$ is unitary then $N_{++}=N_{--}$ for $|p/q|=1$.
\end{theorem}
\noindent \textbf{Proof:}
The first assumption
can be justified by looking at the experimental branching ratios
of the decay channels common to $B^0$ and $\bar B^0$.
Taking $y=0$ leads to $I_{-+}=-I_{+-}$. Using the parameterization 
\be
S = e^{i\phi} \left( \begin{array}{rr} a & b \\ -b^* & a^*
                  \end{array} \right)
\ee
for a general unitary matrix and assuming $p/q = e^{i\delta}$
we find from Eqs. (\ref{int1}) and (\ref{int2})
\be
\int_0^\infty dt  T_{11}^*(t) T_{12}(t) =
- \left( \int_0^\infty dt  T_{22}^*(t) T_{21}(t)  \right)^* =
(I_+ - I_-) a^* b + 
I_{+-} (e^{i\delta} (a^*)^2 + e^{-i\delta} b^2 ) \, .
\ee
Then Eqs. (\ref{++}) and (\ref{--}) provide for
$N_{++}=N_{--}$ under the above conditions. \hfill $\Box$

\medskip

In the following we will stick to the assumptions 
\be \label{ass}
y=0, \quad S \in U(2) \quad \mbox{and} \quad \left| \frac{p}{q} \right|=1 
\ee
for two reasons. First of all, it has
been shown experimentally that the ratio measuring CP violation
in mixing \cite{oku} 
$A_{\mathrm{CP}} \equiv (N_{++}-N_{--})/(N_{++}+N_{--})$
is small \cite{CDF,CLEO}. 
Our assumptions comply with this fact through $A_{\mathrm{CP}}=0$. 
Secondly, the general case, i.e. without the 
assumptions (\ref{ass}) and with
$R$ and $A_{\mathrm{CP}}$ as experimental input, has more
parameters and the conceptual difficulty that we fake CP
violation by the matrix $S$, which we would like to avoid.
We will see in the next section that our simplified
scenario is general enough to illustrate in a clear way the point 
we want to make.

\section{How far is the \bbb\ system from total decoherence?}

Under conditions Eq. (\ref{ass}) the formulas for the number
of dilepton events are very simple. Defining $B \equiv b_+ = b_-$ we
obtain
\ba
N_{++} = N_{--} & = & \frac{B^2}{4\Gamma^2} 
\left\{ \frac{x^2}{1+x^2} + \frac{\zeta Z}{(1+x^2)^2} \right\} \\
N_{+-} = N_{-+} & = & \frac{B^2}{4\Gamma^2}
\left\{ \frac{2+x^2}{1+x^2} - \frac{\zeta Z}{(1+x^2)^2} \right\}
\ea
with
\be \label{ZR}
Z = \left| 2ab^* + ix(e^{-i\delta}a^2+e^{i\delta}(b^*)^2) \right|^2 \, .
\ee
For the ratio of the the number of like-sign dilepton events to
opposite-sign dilepton events we find
\be \label{R}
R = \frac{N_{++}+N_{--}}{N_{+-}+N_{-+}} =
\frac{x^2 + \frac{\zeta Z}{1+x^2}}
{2+ x^2 - \frac{\zeta Z}{1+x^2}} \, .
\ee
For $\zeta = 0$ we have the familiar quantum-mechanical result 
neglecting $y$ and CP violation in \bbb\ mixing.

The relevant properties of $Z$ are summarized in the following
lemma.
\newpage\noindent
\textbf{Lemma:} 
{\it
\begin{enumerate}
\item $Z$ is a function of $S$ and
depends on two variables which can be taken as
$|a|=\sqrt{1-|b|^2}$ and the phase of $e^{-i\delta}ab$. Thus, from the four
parameters of a general unitary 2$\times$2 matrix only two are relevant in
the problem under discussion.
\item The range of $Z$ is given by the interval $0 \leq Z \leq 1+x^2$.
\end{enumerate}
}

\noindent \textbf{Proof:} We parameterize $a$ and $b$ by
\be
a = \cos \omega\, e^{i(\delta + \gamma)}, \quad
b = \sin \omega\, e^{i(\rho -\gamma)}
\ee
and obtain
\be \label{Z}
Z = (\sin 2\omega - x \cos 2\omega \sin \rho)^2 + x^2 \cos^2
\rho \, .
\ee
Thus $Z$ is a function of $\omega$ and 
$\rho = \arg (e^{-i\delta}ab)$, which demonstrates the first
statement. With Eq. (\ref{Z}) we easily check that
\be \label{square}
1 + x^2 - Z = (\cos 2\omega + x \sin 2\omega \sin \rho)^2 \, .
\ee
As a consequence we obtain $Z \leq 1+x^2$. The right-hand side
of Eq. (\ref{square}) is a continuous function of $\omega$ and
$\rho$. Clearly, its minimum is 0 and one can easily show that
the maximum is given by $1+x^2$. Therefore, the range of $Z$ is
specified by the interval in the second statement of the lemma.
\hfill $\Box$

\medskip

The analysis in our previous work Ref. \cite{bg} corresponds to $Z=x^2$ 
whereas the case discussed in Ref. \cite{ds} corresponds to
the maximal value $Z=1+x^2$. Eqs. (\ref{ZR}) and (\ref{R})
represent the main result of this work. The latter formula 
shows that $R$ depends only on the product $\zeta Z$ which is
uniquely determined by measurements of $R$ and $x$. On the other
hand, varying the matrix $S$, though only in the set of unitary
matrices, the function $Z$ varies in the interval given by the
lemma. Clearly, using $R$ of Eq. (\ref{R}) and the experimental
values $\bar R$ and $\bar x$ of $R$ and $x$, respectively, 
for a determination of $\zeta$, the result will strongly depend
on the basis chosen in the $B^0$--$\bar B^0$ space represented
by the matrix $S$. 

In the following we take the values 
$\bar R = 0.189 \pm 0.044$ and $\bar x = 0.74 \pm 0.05$
as given in Ref. \cite{bg}. $\bar R$ has been determined from
the results of the CLEO \cite{CLEO} and ARGUS \cite{ARGUS}
experiments by simply adding the squares of the statistical and
systematic errors for each experiment and using the 
law of combination of errors. The same method has been applied
to determine $\bar x$ from the $\Delta m$ results of the 
four LEP experiments \cite{LEP}
and the $B^0$ lifetime which is found in Ref. \cite{RPP}.

In order to assess the validity of quantum mechanics, there are two
obvious measures associated with the parameter $\zeta$. With $\bar R =
\bar R_0 \pm \Delta \bar R$ and $\bar x = \bar x_0 \pm \Delta \bar x$
as experimental input we can calculate $\bar \zeta = \bar \zeta_0 \pm \Delta
\bar \zeta$. Then the two measures are given by the distance of $\bar \zeta_0$
from 0 and from 1, respectively, each expressed in units of the one standard
deviation $\Delta \bar \zeta$. It is easy to check numerically that 
0 is within the
one standard deviation interval of $\bar \zeta$ 
for the whole range of $Z$ (see also Fig. 1). The
distance of $\bar \zeta_0$ from 1 shows how far the \bbb\ system is from
complete decoherence. In this context we have two extreme
cases. If we chose $S$ such that $Z$ is very close or equal to 0
then in this picture complete decoherence is not excluded. On
the other hand, for $Z=1+x^2$ chosen in Ref. \cite{ds}
complete decoherence is excluded at around 8 standard deviations.
Thus, as was noticed in Ref. \cite{ds}, the question how far the
\bbb\ system is from total decoherence has no unique answer. It
depends on the basis choice represented by the matrix $S$. 
What is significant, however, is the existence of a basis where
the \bbb\ system is far away from total decoherence and where 
the corresponding $\zeta$ is close to 0 in agreement with
quantum mechanics. We have shown that the ``best basis'' in this
respect is given by the mass eigenstates $B_H$, $B_L$ which was
chosen in Ref. \cite{ds}.

In Fig. 1 we have plotted $R$ (Eq. (\ref{R})) as a function of $\zeta$. The
three thin solid lines which are not so steep
represent $R$ with $Z=x^2$ taking into account the
three values $\bar x_0 + \Delta \bar x$ (upper curve), $\bar x_0$
(middle curve) and $\bar x_0 - \Delta \bar x$ (lower curve). The
three steep thin lines are the analogous curves for $Z=1+x^2$. The
thick horizontal lines depict $\bar R$ with its one
standard deviations. For each $Z$ the mean value $\bar\zeta_0$ is found
at the intersections of the middle curves and the middle horizontal
line, whereas $\Delta \bar\zeta$ is approximately given by the
distance of $\bar\zeta_0$ from the $\zeta$ where the lower (upper)
curves cut the upper (lower) horizontal line.
The figure nicely illustrates Eq. (\ref{R}).
As a function of $\zeta$ the ratio $R$ is steeper if $Z$ is
larger. Therefore, for larger $Z$ the allowed range of $\bar\zeta$ is
smaller and closer to 0. We can read off from the figure that the mean
values $\bar\zeta_0$ of $\bar \zeta$ are less than a standard deviation away
from 0 for $Z=x^2$ and $Z=1+x^2$. 
We can also check from the figure that the distance of $\bar\zeta_0$ 
from 1 is nearly 8 standard deviations for the maximal
$Z=1+x^2$. 

\section{Conclusions}

In this paper we have used the ratio 
$R=\,$(\# like-sign dilepton events)/(\# opposite-sign dilepton events)
as an observable testing for interference effects in the \bbb\
state generated by the $\Upsilon (4S)$ decay.
Quantum-mechanically, this \bbb\ state is entangled and extends
over macroscopic distances. On average, $B^0$ and $\bar B^0$ are
separated by $3 \times 10^{-2}$~mm at the time when one of the two particles
decays. We have shown that already with present data for $R$ and
the $B_H$--$B_L$ mass difference interesting conclusions can be
drawn on the decoherence parameter $\zeta$ introduced to
parameterize deviations from the interference term given by
quantum mechanics. Depending on the basis choice in the $B^0$--$\bar B^0$
space and under certain simplifying assumptions, the distance
of the experimentally determined mean value $\bar \zeta_0$
from $\zeta = 1$, which represents total
decoherence, ranges from 0 to 8 standard deviations. The maximal
distance is obtained by using the basis $|B_H \rangle$, $|B_L
\rangle$ of mass eigenstates. 

Though this indefiniteness may
seem confusing at the moment it is merely the consequence of the freedom
in quantum mechanics to choose a basis in the $B^0$--$\bar B^0$
space, which eventually leads to different interference terms for
different basis choices. Parameters $\zeta$ modifying different
interference terms are intrinsically different and adopt
therefore in general different values when confronted with 
experimental data. However, with respect to testing quantum
mechanics it is essential that there exist bases such that the
decoherence parameter it significantly distant from 1 and at
the same time close to 0 \cite{bg,ds}. From this point of view
the presence of an interference term in semileptonic \bbb\ decays
according to quantum mechanics is quite well confirmed and future 
improvement of the
measurements of $R$ and the $B_H$--$B_L$ mass difference allow
to expect even more precise confirmation of interference
effects over macroscopic distances for massive particles.

At present no consistent theory extending quantum mechanics is
known and the introduction of the decoherence parameter might
represent a devious path leading away from consistent extensions.
In any case, this procedure is rather
arbitrary and the analysis of this work corroborates
this impression through the strong basis dependence of the
modification of the quantum-mechanical expressions. Thereby even
CP violation in \bbb\ mixing can be simulated. These features
certainly exist in $K^0 \bar K^0$ and similar
systems as well if they are analogously modified.

\newpage
\begin{figure}
\setlength{\unitlength}{1cm}
\begin{picture}(16,9)
\put(0.8,0.5){\makebox(0,0)[lb]{
\scalebox{1.4}{\includegraphics*[3.5cm,20cm][13cm,26.7cm]{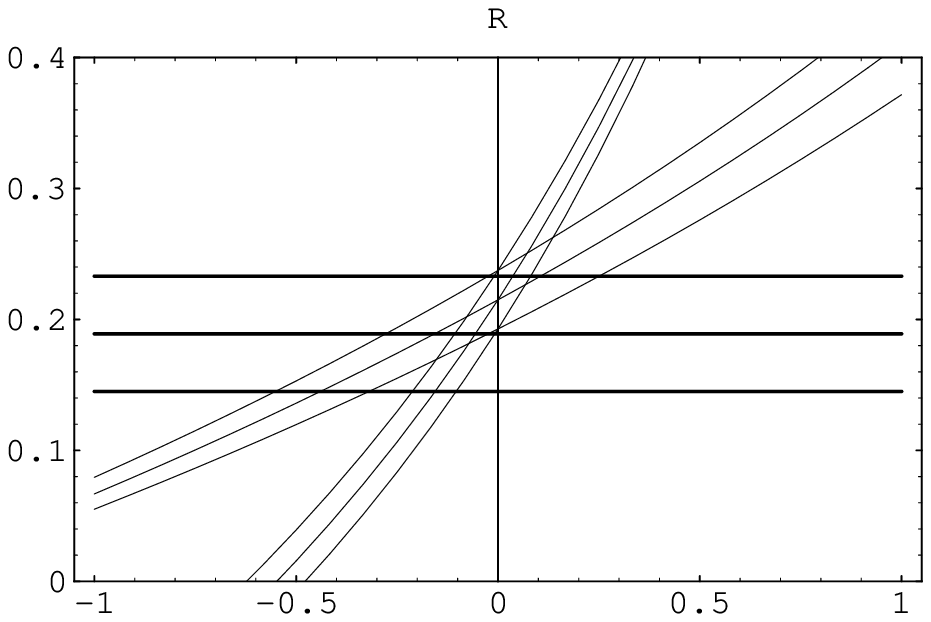}}
}}
\put(14.7,2.2){\makebox(0,0)[l]{$\zeta$}}
\put(8.1,10.2){\makebox(0,0)[b]{$R$}}
\put(10,9.9){\makebox(0,0)[b]{$Z=1+x^2$}}
\put(13.7,9.9){\makebox(0,0)[b]{$Z=x^2$}}
\end{picture}
\caption{The ratio $R$ (Eq. (\ref{R})) as a function of the
decoherence parameter $\zeta$ for $Z=1+x^2$ \cite{ds}, the maximal
$Z$, and $Z=x^2$ \cite{bg}. For each $Z$ the three thin solid lines
correspond to three different values of $x$ given by
$\bar x = 0.74 \pm 0.05$, derived from the experimental
results in Ref. \cite{LEP} and the $B^0$ lifetime taken from
Ref. \cite{RPP}. 
The value of the dilepton ratio $\bar R = 0.189 \pm 0.044$ obtained by
the combined experimental results of Refs. \cite{CLEO} and
\cite{ARGUS} is shown by the thick horizontal lines.}
\end{figure}
\end{document}